\documentclass[12pt]{article}
\usepackage{graphicx}
\begin{document}

\begin{center}

{\large SICS: SOME EXPLANATIONS}

\vspace{10mm}

{\large Ingemar Bengtsson}

\vspace{7mm}

{\sl Stockholms Universitet, AlbaNova \\
Fysikum \\
S-106 91 Stockholm, Sverige}

\vspace{8mm}

{\bf Abstract:}

\end{center}

{\small 

\noindent The problem of constructing maximal equiangular tight frames or SICs was raised 
by Zauner in 1998. Four years ago it was realized that the problem is closely connected to a 
major open problem in number theory. We discuss why such a connection was perhaps to be 
expected, and give a simplified sketch of some developments that have taken place in the 
past four years. The aim, so far unfulfilled, is to prove existence of SICs in an infinite 
sequence of dimensions.}

\vspace{8mm}

{\bf 1. What's in a name?}

\vspace{5mm}

\noindent We will be concerned with configurations of vectors known as SICs, and 
pronounced `seeks' because a proof of their existence in all finite dimensions 
is being sought \cite{Chris}. The problem is easy to state, but soon reveals unexpected 
depths. A little more generally we want to find sets of $N$ unit vectors 
in the finite dimensional Hilbert space ${\bf C}^d$, and constants $c_1$ and $c_2$, 
such that 

\begin{equation} \sum_{i=1}^N|\psi_i\rangle \langle \psi_i| = c_1 {\bf 1} \label{tightframe} 
\end{equation}

\begin{equation} |\langle \psi_i|\psi_j\rangle |^2 = c_2 \hspace{5mm} \mbox{if} 
\hspace{5mm} i \neq j \ . \end{equation}

\noindent Such sets are called {\it equiangular tight frames} \cite{Waldron}. They can 
be thought of as $N$ equidistant points in complex projective space, or as a regular 
simplex in the space containing the convex set of all mixed quantum states, carefully 
centred and arranged so that all its corners are pure. One proves easily that if the 
arrangement can be done at all then 

\begin{equation} d \leq N \leq d^2 \ , \hspace{10mm} 
c_1 = \frac{N}{d} \ , \hspace{10mm} c_2 = \frac{N-d}{d(N-1)} \ . \end{equation}

\noindent Existence is not a foregone conclusion. If $d = 3$ the possible values of 
$N$ are 3, 4, 6, 7, and 9, while $N = 5$ and $N = 8$ are impossible \cite{Feri}. 
Minimal ETFs are known as {\it orthonormal bases}, while maximal ETFs consisting 
of $d^2$ unit vectors are called {\it SICs}. At first 
they were called {\it Maximale Quantendesigns} \cite{Zauner}. Finding SICs is 
of interest in classical signal processing and in quantum information theory. 
In the latter context the long acronym {\it SIC-POVM} is often used, and then the 
first three letters stand for ``symmetric informationally complete'' and the last 
four for ``positive operator valued measure'' \cite{Renes}. Fortunately, in some 
quantum applications there are conceptual reasons to drop the ungainly last set 
of four letters \cite{Tavakoli}. 

SICs have a background in engineering, but they have recently moved into unexplored 
regions of algebraic number theory. In 2016 it was conjectured that the numbers needed to 
construct them are the kind of numbers that appear in the first unsolved case of Hilbert's 
12th problem \cite{AFMY, AFMY2}. The hypothesis was supported by some 
solid evidence \cite{Scott, ACFW}. It bears the hall- mark of 
truth, because over the last four years it has led to explanations, and predictions, 
of a very large number of bewildering facts about SICs in various dimensions. Thus 
the status of the SIC existence problem has changed. There always were good reasons 
to seek them \cite{FHS}, but now they also seem to be intimately connected to a grand 
unsolved problem in number theory. In the spirit of the V\"axj\"o meetings 
\cite{Vaxjo} we hope to provide at least some 
explanations of this development here: Of the way that the connection to number 
theory arises (Section 2), of the finite groups that generate SICs and their 
symmetries (Section 3), and of how the theory as developed so far organizes Hilbert 
spaces of different dimensions into sequences (Section 4). To keep the discussion 
simple we will restrict the technical part to the case of odd dimensions only. We do 
this with some regret because, like the rotation group \cite{Kreisel}, the groups that 
generate SICs treat even dimensions in a subtle but ultimately very satisfying way. 

There is a school of thought maintaining that SICs will ultimately prove to be as 
important \cite{Chris, qplex, QBism} for quantum foundations as are the orthonormal bases 
\cite{Gleason}. We do not pursue this argument here, but we do hope to convince the 
reader that SICs deserve to be spelt with capital letters.

\vspace{10mm}

{\bf 2. Smelling the problem}

\vspace{5mm}

\begin{figure}[ht]
\begin{center}
\includegraphics[width=90mm]{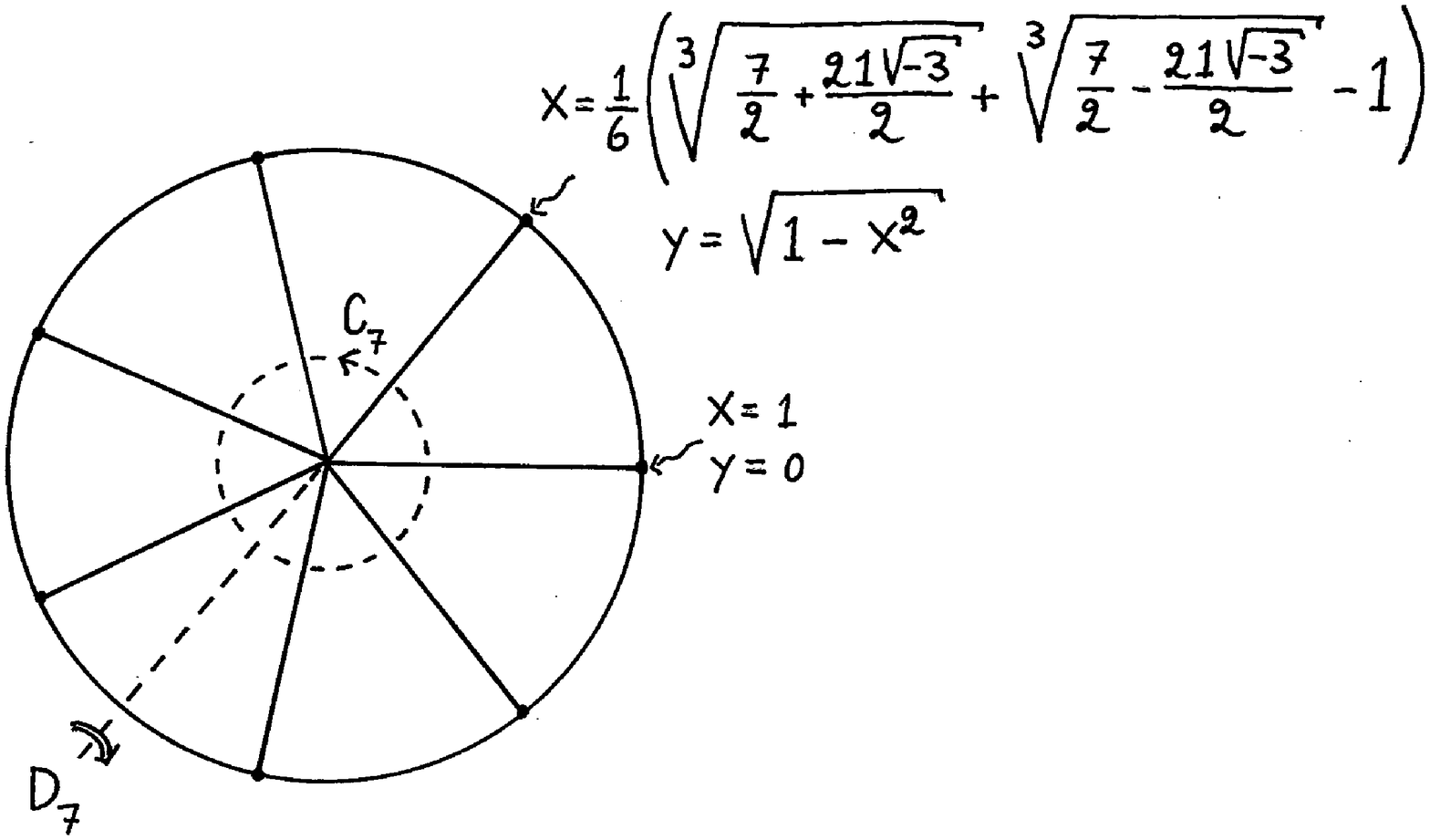}
\end{center}
\caption{\small The construction of a regular heptagon involves both group 
theory and number theory. Incidentally, the fact that cube roots appear in 
the numbers means that the heptagon cannot be constructed with ruler and 
compasses.}
\label{fig:heptagon}
\end{figure}

\noindent Why are SICs so hard to find, and why are they connected to number 
theory? To see this we begin with the famous problem of dividing a circle into $n$ 
equal parts. We choose $n = 7$ as our example. Some group theory clearly enters the 
problem. What we need are the seven corners of a regular heptagon inscribed in the 
circle, and we observe that these corners are the orbit of an abelian and cyclic 
group, $C_7$. On closer inspection we realize that the heptagon is left invariant 
by a larger dihedral group, which is conveniently thought of 
as a subgroup of the rotation group $SO(3)$. Let us choose one corner to sit at 
$(x,y) = (1,0)$. As it turns out, placing the remaining six corners leads us to 
perform some complicated root extractions in order to find 
their coordinates $x$ and $y$. See the illustration in Figure \ref{fig:heptagon}. 

If we view the plane as a complex plane the position 
of the corner in the Figure \ref{fig:heptagon} is given by the complex number 
$\omega = x+iy$. It must be a seventh root of unity, so that $\omega^7 = 1$. 
It follows that the coordinates of the six corners that we need to construct are 
the six roots of the polynomial 

\begin{equation} p(z) = z^6 + z^5 + z^4 + z^3 + z^2 + z + 1 \ . \label{om7} \end{equation}

\noindent There must be something special about our polynomial, because the roots 
of a generic polynomial of degree higher than four cannot be given by root extractions 
at all. This point was explained by Galois, who considered the group that permutes 
the roots of a given polynomial equation. In our case this group is easily identified. 
Suppose we consider the permutation that takes $\omega$ to $\omega^3$. For consistency 
we can then deduce that 

\begin{equation} \omega \rightarrow \omega^3 \rightarrow \omega^9 = \omega^2 
\rightarrow \omega^6 \rightarrow \omega^{18} = \omega^4 \rightarrow \omega^{12} 
= \omega^5 \rightarrow \omega^{15} = \omega \ . \end{equation}

\noindent We have gone through all the roots! It follows that the {\it Galois group} 
of the polynomial is the abelian group $C_6$, and Galois proved that roots of 
polynomials having an abelian Galois group can always be given in terms of root 
extractions \cite{Stewart}. 

Root extraction is a complicated affair, but a beautiful and extremely important 
feature is waiting in the wings. Introduce the transcendental function 

\begin{equation} e(x) = e^{2\pi i x} \ . \end{equation}

\noindent We get the seven corners of our heptagon by evaluating this function 
at seven rational points. And we see that the trigonometric and exponential 
functions---discoveries without which our modern civilization would be 
unthinkable---appear naturally out of the regular polygon problem. 

When the problem of the regular $n$-gon is viewed with the eyes of number 
theory, the key role is played by the {\it cyclotomic} or circle-dividing 
number field with {\it conductor} $n$. We obtain this number field by adding an 
$n$th root of unity to the set of rational numbers and then applying addition 
and multiplication in every possible way to this set, keeping in mind that 
$\omega_n$ solves an equation analogous to that of setting the polynomial 
(\ref{om7}) to zero. In this way we take the step from ${\bf Q}$ to 
${\bf Q}(\omega_n)$. The extended number field can be described as a 
vector space over ${\bf Q}$. For $n = 7$ this vector space has 6 dimensions, 
because any number in ${\bf Q}(\omega_7)$ is a linear combination of the basis 
vectors $1, \omega, \dots , \omega^5$. The dimension of the vector space is 
also known as the {\it degree} of the number field. 

Kronecker and Weber proved that every 
abelian extension of the rational field, that is to say every extension whose Galois 
group over the rationals is an abelian group, is a subfield of a cyclotomic field 
with some conductor $n$. Given that we have an elegant description of the generators 
of all the cyclotomic fields in terms of a transcendental function evaluated at 
rational points, this is very satisfactory. A generator of the cyclotomic 
field with conductor $n$ is obtained as $e(1/n) = e^{2\pi i/n}$. We use the notation 
$\omega_n = e^{2\pi i/n}$ from now on. 

Then Kronecker had a dream. Start with the degree 2 number field obtained by adding 
the square root of an integer $D$ to ${\bf Q}$, and consider the most 
general abelian extension of that number field. The Galois group of this field 
considered as an extension of ${\bf Q}$ need not be abelian, but it will enjoy 
a {\it normal series} of the form $e \triangleleft H \triangleleft G$, where $H$ is the Galois 
group of ${\bf Q}(\sqrt{D})$ over ${\bf Q}$. The notation means that $H$ is 
a normal subgroup of $G$ and that the group $G/H$ is abelian. Hence the Galois 
group is close to abelian, and again Galois assures us that the numbers that 
occur can be arrived at from the rationals by root extractions. Kronecker saw, 
as in a vision, that if the quadratic base field we start out with uses a negative 
integer $D$, then it should be possible to complete the story to find special 
{\it ray class fields} housing every abelian extension of this base field, and 
moreover it should be possible to find transcendental functions (elliptic and 
modular, in this case) such that the generators of the ray class fields are 
obtained by evaluating the functions at special points on special elliptic curves. 

The program was not quite completed by the time Hilbert posed his famous problems 
for the twentieth century, but it was soon after \cite{comedy}. Hilbert was impressed. 
In his 12th problem he asked for a similar description of the most general abelian 
extension of an arbitrary base field \cite{Hilbert}. Mathematicians soon set to work on 
this, wisely concentrating on the simplest open case, that of {\it real} quadratic base 
fields ${\bf Q}(\sqrt{D})$ with $D > 0$. The twentieth century proved too short 
for the task. Still a classification of the relevant ray class fields was achieved. 
They are specified by two positive integers, $D$ that gives the base field and 
$d$ which gives the conductor. Algorithms for finding generators of these ray 
class fields have been implemented in computer algebra packages. But 
the problem of writing down explicit generators for them, preferably by starting 
from some transcendental function, remains open. As the example of the heptagon shows, 
a solution may have far-reaching consequences. 

Now we come to the point. The Ray Class Hypothesis states that the numbers needed to 
construct a SIC in dimension $d > 3$ generate a ray class field over a real quadratic 
base field \cite{AFMY, AFMY2}. The conductor of the field is equal to $d$ if $d$ is 
odd and $2d$ if $d$ is even, while the integer $D$ that determines the base field is 
given by \cite{AYAZ}

\begin{equation} D = (d+1)(d-3) \ . \label{AYAZd} \end{equation}

\noindent We will return to this interesting formula in Section 4. It does 
make it appear as if the SICs may be the geometrical objects that hold the 
key to a part of Hilbert's 12th problem. 

There is a complication here, which is that in almost all of the dimensions that have 
been investigated several unitarily inequivalent SICs do exist \cite{Scott, Andrew}. 
Then the hypothesis says that at least one of them can be constructed using the ray 
class field. If one of the SICs is singled out as having the highest symmetry, this 
is it \cite{ACFW}. The other SICs require further abelian extensions of the ray class field. 

As a preparation for Section 3 we notice the fact that for the special choice of 
conductors implied by the formula the ray class field will contain the cyclotomic 
field ${\bf Q}(\omega_{2d})$ as a subfield \cite{AFMY}. We also note that if $d$ 
is odd then $-\omega_d$ is a $2d$th root of unity, and it certainly belongs 
to ${\bf Q}(\omega_d)$. Thus ${\bf Q}(\omega_d) = {\bf Q}(\omega_{2d})$ when $d$ 
is odd, but not when $d$ is even. As a preparation for Section 
4 we observe that if we fix the base field and consider two different conductors 
$d_1$ and $d_2$ then, by the very definition of conductors, the ray class field 
with conductor $d_1$ is a subfield of that with conductor $d_2$ if and only if 
$d_1$ is a divisor of $d_2$. The reader may easily check this for the special 
case of cyclotomic fields. 

\vspace{10mm}

{\bf 3. The acting groups}

\vspace{5mm}

\noindent To find SICs we first ask if they are orbits of a group, as the 
$n$-gons are. Gerhard Zauner, and independently Joseph Renes and his coworkers, 
conjectured that a discrete Heisenberg group plays this role \cite{Zauner, Renes}. 
This group is a central extension of the product 
of two cyclic groups $C_d\times C_d$, and its unitary representation is 
essentially unique. Its unitary automorphism group, from which extra symmetries 
of the SICs are taken, contains as a factor group the discrete symplectic group 
acting on the discrete `plane' of the group elements. Zauner made a further 
mysterious conjecture, later sharpened \cite{Marcus} to say that every SIC has 
an extra symmetry of order 3. Closer examination led to more detailed 
conjectures about higher symmetries appearing in special cases in special dimensions 
\cite{Scott, Andrew, GS}. 

Numerical searches for SICs are made easier once it is assumed that they are 
orbits under a group. It is then enough to find a single {\it fiducial vector} 
from which the group creates the SIC. In this ´way SICs have been found numerically 
in all dimension $d \leq 193$ and in some higher dimensions, the record being $d =2208$ 
\cite{Andrew, MGun}. In his thesis Zauner also provided exact solutions in dimensions 
4 and 5. (Dimension 2 is trivial. A solution in dimension 3, related to an elliptic 
curve invariant under this very group, was provided by Hesse in 1844 \cite{Hesse}.) 
By now more than one hundred exact solutions are known \cite{Scott, ACFW, MGun}. 

Heisenberg groups are important throughout quantum mechanics and signal processing 
alike. For us a convenient starting point is the book by Weyl \cite{Weyl}. The 
Weyl--Heisenberg group is generated by $X, Z, \omega$, subject to the 
relations 

\begin{equation} ZX = \omega XZ \ , \hspace{5mm} \omega X = X\omega \ , \hspace{5mm} 
\omega Z = \omega Z \ , \hspace{5mm} X^d = Z^d = \omega^d = {\bf 1} \ .
\end{equation}

\noindent There is one such group for each choice of the integer $d$. Weyl thought of 
them as toy models of the group that encapsulates the non-commutativity of position 
and momentum, which at some point became known as the Heisenberg group. The discrete 
group has an essentially unique irreducible 
unitary representations in a Hilbert space of dimension $d$. We first fix $\omega$ to be 

\begin{equation} \omega = \omega_d = e^{\frac{2i \pi}{d}} \end{equation}

\noindent (times the unit matrix, which is understood). Actually any primitive $d$th 
root of unity would do, which is why the representation is only `essentially' unique. 
In the basis where $Z$ is chosen to be diagonal it is 

\begin{equation} Z|i\rangle = \omega^i|i\rangle \ , \hspace{8mm} X|i\rangle = 
|i+1\rangle \ , \label{clockandshift} \end{equation}

\noindent where the basis vectors are labelled by integers counted modulo $d$. 

The representation makes use only of numbers from the cyclotomic field 
${\bf Q}(\omega_d)$. But if the dimension $d$ is odd then ${\bf Q}(\omega_d) = 
{\bf Q}(\omega_{2d})$. To save the one-to-one correspondence between dimensions 
and cyclotomic fields one can extend the centre of the group to include $2d$th 
roots of unity if $d$ is even. There are actually several good reasons for this 
move \cite{AFMY, Marcus}, but as advertized in the introduction we will restrict 
the discussion to the case of odd $d$ from now on in order to keep the story brief. 

To understand how the Weyl--Heisenberg group depends on the dimension $d$ we 
begin by recalling an interesting fact about cyclic groups. It is easy to see 
that $C_4 \neq C_2\times C_2$, because every element of the product group 
squares to the identity, while $C_4$ contains elements of order 4. On the other 
hand it is easy to convince oneself that $C_6 = C_2\times C_3$. What makes this 
case different is that 2 and 3 are {\it relatively prime}, that is to say their 
greatest common divisor equals 1. Here it means that the cyclic 
groups, and indeed the Weyl--Heisenberg groups, can be broken down into relatively 
prime atoms. That is to say, if $H(d)$ denotes the group in dimension $d$, 
and if $d$ can be decomposed into primes as 

\begin{equation} d = p_1^{n_1}\cdot p_2^{n_2}\cdot \dots \cdot p_r^{n_r} \ 
\end{equation}

\noindent then 

\begin{equation} H(d) = H(p_1^{n_1})\times H(p_2^{n_2}) \times \dots \times 
H(p_r^{n_r}) \ . \end{equation}

\noindent To prove this one applies the Chinese remainder theorem from elementary 
number theory \cite{Hardy}. So it suffices to understand the group in prime power 
dimensions. We remark 
that if the dimension is prime then it can be proved that every SIC that is generated by 
a group is generated by the Weyl--Heisenberg group \cite{Huangjun}. 

The central extension, with its troublesome phase factors, is there only to provide 
an interesting representation theory. When the group acts on projective space it does 
so as the product of two cyclic groups, so it makes sense to select a set of only 
$d^2$ group elements to work with. Nevertheless it pays to pay careful attention to 
the phase factors when we do so. We define the {\it displacement operators} \cite{Marcus} 

\begin{equation} D_{i,j} = \tau^{ij}X^iZ^j \ \ , \hspace{8mm} \tau = - e^\frac{\pi i}{d} \ 
. \end{equation}

\noindent Notice that $\tau$ is a $d$th root of unity when $d$ is odd, as we assumed 
it to be for simplicity. A key fact about the displacement operators is that they 
form a unitary operator basis, which is why Schwinger allowed the Weyl--Heisenberg 
group to take the centre stage in his presentation of quantum mechanics \cite{Schwinger}. 
The group law takes the form 

\begin{equation} D_{i,j}D_{k,l} = \tau^{ki-lj}D_{i+k,j+l} \hspace{5mm} 
\Leftrightarrow \hspace{5mm} D_{\bf p} D_{\bf q} = \tau^{<{\bf p},{\bf q}>}
D_{{\bf p} + {\bf q}} \ . \label{grlaw} \end{equation}

\noindent Here we introduced two-component `vectors' ${\bf p}, {\bf q}$ whose 
components are integers modulo $d$, as well as the symplectic form $<{\bf p},{\bf q}>$. 
The latter is very useful when we consider the unitary automorphism group, that is 
to say the group of unitary operators that permute the elements of the 
Weyl--Heisenberg group under conjugation. It is known as the {\it Clifford group} 
\cite{Bolt}, and contains the symplectic group $SL(2,{\bf Z}_{d})$ as a factor group. 
The latter has a defining representation in terms of two-by-matrices $F$ obeying 

\begin{equation} < F{\bf p}, F{\bf q}> \ = \ <{\bf p}, {\bf q}> \ {\rm modulo} \ d 
\ . \end{equation}

\noindent These are precisely the matrices having unit 
determinant and entries that are integers modulo $d$. Once the representation of the 
Weyl--Heisenberg has been chosen the unitary representative $U_F$ of the symplectic 
matrix $F$ is completely fixed up to phase factors \cite{Marcus} by the 
defining relation 

\begin{equation} U_FD_{\bf p}U_F^{-1} = D_{F{\bf p}} \ . \end{equation}
 
\noindent Here we want to stress that the entire Clifford group is represented by 
matrices all of whose entries lie in the cyclotomic number field. Acting on any vector 
whose components are built using a number field that includes the cyclotomic field, 
it will produce new vectors built from the same kind of numbers. This is clearly 
relevant for us. 

The Chinese remainder theorem again makes itself felt at this point, 
so that the Clifford group splits into a direct product determined by the 
decomposition of the dimension into prime factors: it is enough to 
understand how it behaves in prime power dimensions. For $ d= 3, 5$ the symplectic 
groups enjoy the group isomorphisms 

\begin{equation} SL(2,{\bf Z}_3)/\pm {\bf 1} = {\bf T} \ \ \end{equation}

\begin{equation} SL(2,{\bf Z}_5)/\pm {\bf 1} = {\bf I} \ , \end{equation}

\noindent where ${\bf T}$ and ${\bf I}$ are the symmetry groups of the tetrahedron 
and the dodecahedron (or icosahedron), respectively. A reader equipped with cardboard 
models of these polyhedra can therefore take in the structure of these groups at a 
glance. But some structure is hidden since we have divided out the centre of the 
symplectic group. It consists of an order two matrix with a simple unitary representative, 

\begin{equation} 
F = - {\bf 1} \hspace{5mm} \Rightarrow \hspace{5mm} \langle i|U_F|j\rangle = 
\delta_{0,i+j} \ . \label{parity} \end{equation} 

\noindent We denote this particular unitary operator $U_F$ either as $U_P$ or 
as $A_{\bf 0}$. It is known as the {\it parity operator}. We can use it to construct 
an alternative unitary and Hermitean 
operator basis, consisting of the {\it phase point operators} \cite{Wootters}

\begin{equation} A_{\bf p} = D_{\bf p}A_{\bf 0}D_{\bf p}^{-1} \ . \end{equation}

\noindent This operator basis is significant in the SIC problem in more ways 
than one. One can show that the 
spectrum of a phase point operator consists of $(d+1)/2$ eigenvalues $+1$, and $(d-1)/2$ 
eigenvalues $-1$. (Our self-imposed restriction to odd $d$ is still in force.) It 
follows that the phase point operators give us a set of $d^2$ subspaces of dimension 
$(d+1)/2$, defined by the projectors 

\begin{equation} \Pi_{\bf p} = \frac{1}{2}({\bf 1} + A_{\bf p}) \ . \end{equation}

\noindent We can think of these subspaces as points in the Grassmannian 
$Gr_{(d+1)/2, d}$, analogously to how we regard one-dimensional subspaces as points 
in projective space. There is a natural notion of chordal distance $D_{\rm ch}$ 
between points in a Grassmannian, which if we identify the subspaces with the 
projectors is given by 

\begin{equation} D_{\rm ch}^2(\Pi_{\bf p}, \Pi_{\bf q}) = \mbox{Tr}(\Pi_{\bf p} 
- \Pi_{\bf q})^2 \ . \label{chordal} \end{equation}

\noindent A quick calculation confirms that the subspaces defined by the operator 
basis form a set of $d^2$ equidistant points in the Grassmannian. Exactly why 
we bring up this curious point will become clear in Section 4, but it may not 
be amiss to remark that these subspaces play a role in the theory of elliptic normal 
curves transforming into themselves under the Weyl--Heisenberg group \cite{Hulek}. 

\vspace{10mm}

{\bf 4. To build a ladder to the stars}

\vspace{5mm}

\noindent We now return to the key formula (\ref{AYAZd}), to see how different 
dimensions are connected to each other by the number theoretical properties of 
the SICs they contain. We rewrite it a little by setting $D = m^2D_0$, where 
$m$ is any integer and $D_0$ does not have square factors. Clearly 
${\bf Q}(\sqrt{D}) = {\bf Q}(\sqrt{D_0})$. The formula becomes 

\begin{equation} (d+1)(d-3) = D = m^2D_0 \ . \end{equation}

\noindent It can be read in two directions. If we begin in a Hilbert space of 
dimension $d$ we use it to determine $D$, and hence the base field needed in 
the construction of SICs. But we can also fix the square-free part $D_0$ and 
solve the Diophantine equation for $d$ in order to establish, via the SICs, a 
number theoretical connection between different dimensions. Because the integer 
$m$ is free the result is an infinite sequence $\{ d_i\}_{i=1}^\infty$ of 
dimensions known as a {\it dimension tower}. The entries of the sequence are 
given by a simple formula \cite{AFMY, AFMY2} which however we do not give here 
since this would require a detour to introduce the unit group of the base field 
\cite{Hardy}. 
Instead we give the beginnings of two such towers in Figure \ref{fig:stegar}. 

{\small
\begin{figure}[t]
\begin{picture}(350,180)
\put(70,5){$d={\bf 4}$}
\put(80,21){{\bf 8}}
\put(147,35){{\bf 19}}
\put(76,50){{\bf 48}}
\put(110,65){{\bf 124}}
\put(144,80){{\bf 323}}
\put(39,95){{\bf 844}}
\put(72,105){2208}
\put(1,118){5779}
\put(103,135){15128}
\put(83,15){\line(0,1){4}}
\put(83,29){\line(0,1){18}}
\put(83,57){\line(0,1){45}}
\put(83,112){\line(0,1){54}}
\put(153,88){\line(0,1){80}}
\put(153,43){\line(0,1){33}}
\put(118,73){\line(0,1){59}}
\put(117,144){\line(0,1){23}}
\put(48,103){\line(0,1){63}}
\put(13,127){\line(0,1){40}}
\put(88,14){\line(1,2){25}}
\put(87,28){\line(1,4){26}}
\put(79,15){\line(-1,3){26}}
\put(274,5){$d={\bf 5}$}
\put(279,21){{\bf 15}}
\put(352,35){{\bf 53}}
\put(275,51){{\bf 195}}
\put(315,65){725}
\put(347,80){2703}
\put(239,95){10085}
\put(273,105){37635}
\put(201,117){140453}
\put(306,135){524175}
\put(285,15){\line(0,1){4}}
\put(285,29){\line(0,1){18}}
\put(285,59){\line(0,1){44}}
\put(285,112){\line(0,1){55}}
\put(358,43){\line(0,1){34}}
\put(358,89){\line(0,1){78}}
\put(322,74){\line(0,1){59}}
\put(322,146){\line(0,1){21}}
\put(253,103){\line(0,1){64}}
\put(218,128){\line(0,1){39}}
\put(291,14){\line(1,2){24}}
\put(290,29){\line(1,4){26}}
\put(280,15){\line(-1,3){26}}
\put(185,0){\line(0,1){172}}
\put(186,0){\line(0,1){172}}
\put(5,15){$D_0=5$}
\put(205,15){$D_0=3$}
\end{picture}
\caption{{\small The first 10 dimensions connected by $D_0 = 5$ and by $D_0= 3$. Dimensions 
for which exact SICs are known \cite{Scott, GS, MGun, SiSi} are in boldface. A field is a 
subfield of another if its conductor divides that of the other. In the picture this happens 
if the other field can be reached by walking along upwards directed links. When 
$d$ is even the conductor equals $2d$, but this does not affect this ordering. 
Vertical lines (or `ladders') arise from the substitution $d\rightarrow d(d-2)$.}}
\label{fig:stegar}
\end{figure}
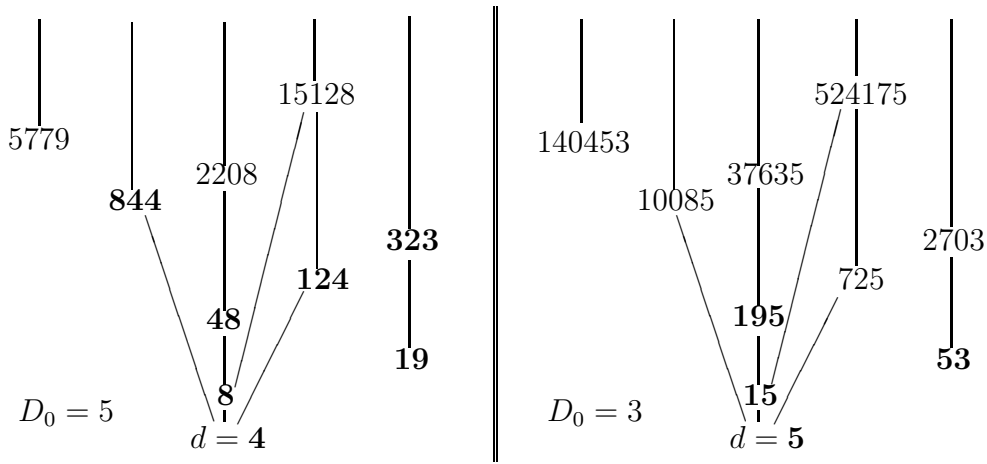
}

The Figure encodes information about when an entry $d_i$ is a divisor 
of another entry $d_j$. This is important because when one conductor is a divisor of 
another it implies that the first ray class field is contained in the other. The 
vertical {\it ladders} in Figure \ref{fig:stegar} are meant to attract attention. 
They arise from the simple observation that 

\begin{equation} d \rightarrow d(d-2) \hspace{5mm} \Rightarrow \hspace{5mm} 
(d+1)(d-3) \rightarrow (d-1)^2(d+1)(d-3) \ . \end{equation}

\noindent The square free part $D_0$, and hence the base field, is unchanged by 
the substitution. Moreover, since (obviously) $d$ divides $d(d-2)$ the field used 
in the higher dimension always contains that used in the lower. With our choice of 
labelling, every odd numbered entry $d_{2n+1}$ in a tower starts its own ladder, 
while the entries $d_{2^r\cdot (2n+1)}$ are said to sit on rung $r$ of some ladder. 

It is tempting to believe that some unknown physics is hidden in these sequences. 
It is also tempting to believe that one can prove SIC existence for all the dimensions 
in such a sequence in some inductive way. At the moment this is just a dream, but 
bits and pieces of what looks like an argument to this effect have materialized 
\cite{aligned, AD, tight, Kiev}. We can simplify the story quite a bit by starting 
it from an observation by Renes et al. \cite{Renes}, and this is what we propose to 
do here. 

Let $|\psi_0\rangle$ be a fiducial vector for a SIC in dimension $d$. Form the vector 
$|\psi_0\rangle \otimes |\psi_0\rangle$ in the symmetric subspace 
${\bf C}^{d(d+1)/2}$ of ${\bf C}^{d^2}$. The Weyl--Heisenberg group can be 
made to act on this vector. Pausing to polish our conventions we define $\omega = 
e^{2\pi i/d}$ and 

\begin{equation} \left. \begin{array}{ccc} {\bf X} & = &  X\otimes X \\ {\bf Z} & = & 
Z^\frac{d+1}{2}\otimes Z^\frac{d+1}{2} \end{array} \right\} 
\hspace{5mm} \Rightarrow \hspace{5mm} {\bf Z}{\bf X} = \omega {\bf X} {\bf Z} 
\ . \end{equation}

\noindent Note that our restriction to odd $d$ is still in force. The case of even 
$d$ is more interesting \cite{AD, Kiev}, but takes more space. We now have the Weyl--Heisenberg 
group $H(d)$ acting on ${\bf C}^{d^2}$, and can define the displacement operators 

\begin{equation} \tilde{D}_{ij} = \tau^{ij}{\bf X}^i{\bf Z}^j \ . \end{equation}

\noindent By assumption 

\begin{equation} \langle \psi_0|D_{\bf p}|\psi_0\rangle = \left\{ \begin{array}{ccl} 1 & 
\mbox{if} & {\bf p} = {\bf 0} \\ \frac{e^{i\theta_{i,j}}}{\sqrt{d+1}} & \mbox{if} & {\bf p} \neq 
{\bf 0} \ , \end{array} \right. \end{equation} 

\noindent where the phase factor is a {\it SIC overlap phase} from dimension $d$. Using 
the definition of $\tilde{D}_{\bf p}$ we see that 

\begin{equation}  \langle \psi_0|\langle \psi_0|\tilde{D}_{\bf p}|\psi_0\rangle 
|\psi_0\rangle = \left\{ \begin{array}{ccl} 1 & \mbox{if} & {\bf p} = {\bf 0} \\ 
\frac{e^{2i\theta_{i,j'}}}{d+1} 
& \mbox{if} & {\bf p} \neq {\bf 0} \end{array} \right. \ , 
\hspace{5mm} j' = \frac{d+1}{2}j \ . \end{equation} 

\noindent From the SIC in dimension $d$ we have obtained an ETF consisting of $d^2$ vectors 
in dimension $d(d+1)/2$ \cite{Renes}. 

We are representing the Weyl--Heisenberg group $H(d)$ in ${\bf C}^{d^2}$, and from Weyl's 
book we know that the representation must be reducible \cite{Weyl}. To take advantage of 
this we introduce the orthonormal basis 

\begin{equation} |ii\rangle = |i\rangle |i\rangle , \hspace{2mm} |(i,j)\rangle 
= \frac{1}{\sqrt{2}}(|i\rangle |j\rangle + |j\rangle |i\rangle ) , \hspace{2mm} 
|[i,j]\rangle = \frac{1}{\sqrt{2}}(|i\rangle |j\rangle - |j\rangle |i\rangle ) \ . 
\end{equation} 

\noindent We now forget about the tensor product structure, and introduce a new one. 
With a suitable ordering of the new basis vectors we can ensure that the representation 
uses block diagonal matrices $\tilde{D}_{\bf p}$ carrying copies of the dimension $d$ 
displacement operators $D_{\bf p}$ in the blocks. Hence we can write  

\begin{equation} \tilde{D}_{\bf p} = {\bf 1}\otimes D_{\bf p} \ , \end{equation}

\noindent where we let the dimension in which the identity operator acts depend on the 
context. If we act on ${\bf C}^{nd}$ we need $n$ blocks, and the dimension we need is $n$. 

Elementary linear algebra tells us if there exists an ETF with $d^2$ vectors in 
dimension $d(d+1)/2$ then there must exist an ETF with $d^2$ vectors 
in dimension $d(d-1)/2$. To see why we renormalize the vectors by defining

\begin{equation} {\bf u}_{\bf p} = \sqrt{d+1}\tilde{D}_{\bf p} |\psi_0\rangle |\psi_0\rangle 
\ . \end{equation}

\noindent We let these vectors form the columns of a $d(d+1)/2\times d^2$ matrix $M$. 
From the tight frame condition (\ref{tightframe}) it follows that the rows of this 
matrix are orthogonal to each other, $MM^\dagger = 2d{\bf 1}_{d(d+1)/2}$. We then 
fill out this rectangular matrix to a unitary matrix

\begin{equation} U = \frac{1}{\sqrt{2d}}\left[ \begin{array}{cccc} {\bf u}_0 & {\bf u}_1 & 
\dots & {\bf u}_{d^2-1} \\ \hline {\bf v}_0 & {\bf v}_1 & \dots & {\bf v}_{d^2-1} 
\end{array} \right] \ , \hspace{8mm} UU^\dagger = U^\dagger U = {\bf 1}_{d^2} \ . 
\label{unitary} \end{equation}

\noindent This is always possible. From unitarity it follows that 

\begin{equation}  ({\bf v}_{\bf p},{\bf v}_{\bf p})  = 
\left\{ \begin{array}{ccl} d-1 & \mbox{if} & {\bf p} = {\bf 0} \\ - e^{2i\theta_{i,j'}} 
& \mbox{if} & {\bf p} \neq {\bf 0} \ . \end{array} \right. \end{equation} 

\noindent Finally we renormalize the ${\bf v}_{\bf p}$ to obtain a set of unit 
vectors in ${\bf C}^{d(d-1)/2}$, and we sneak in the assumption that the 
columns of $U$ are generated, in their entirety, by acting with $\tilde{D}_{\bf p}$ on 
the first column. We obtain 

\begin{equation} |\Psi_{\bf p}\rangle = \frac{1}{\sqrt{d-1}}{\bf v}_{\bf p} = 
\tilde{D}_{\bf p}|\Psi_0\rangle \ . \end{equation} 

\noindent We now have an equiangular tight frame in ${\bf C}^{d(d-1)/2}$, 

\begin{equation}  \langle \Psi_0|\tilde{D}_{\bf p}|\Psi_0\rangle  = 
\left\{ \begin{array}{ccl} 1 & \mbox{if} & {\bf p} = {\bf 0} \\ - 
\frac{e^{2i\theta_{i,j'}}}{d-1} & 
\mbox{if} & {\bf p} \neq {\bf 0} \ . \end{array} \right. \end{equation} 

\noindent This is known as the {\it Naimark complement} of the ETF we started 
out with. We know that it exists, and its Gram matrix is completely known. 

A little rewriting will reveal what we are driving at: 

\begin{equation} \langle \Psi_0|\tilde{D}_{\bf p}|\Psi_0\rangle  = - 
\frac{e^{2i\theta_{i,j'}}}{d-1} \hspace{2mm} \Leftrightarrow \hspace{2mm} 
\langle \Psi_0|{\bf 1}_{\frac{d-1}{2}}\otimes D^{(d)}_{\bf p}|\Psi_0\rangle  = - 
\frac{e^{2i\theta_{i,j'}}}{\sqrt{d(d-2) + 1}} \ . \end{equation}

\noindent The displacement operators that occur here are those for dimension 
$d$, but the absolute value of the right hand side is that appropriate for a SIC 
in dimension $d(d-2)$, the dimension one rung above the dimension we started 
out with. The vectors do not sit in that dimension, but we now observe that 

\begin{equation} \frac{d-1}{2} = \frac{d-2 + 1}{2} \ . \end{equation}

\noindent From Section 3 we recall that this is the dimension of the positive 
parity eigenspace in ${\bf C}^{d-2}$. Hence we can embed the fiducial vector 
$|\Psi_0\rangle$ in that eigenspace to obtain a vector in ${\bf C}^{d-2}
\otimes {\bf C}^d$, taking care to adjust the basis so that the representation 
of the parity operator $U_P$ becomes the standard one (\ref{parity}). We then have 

\begin{equation} \langle \Psi_0|{\bf 1}_{d-2}\otimes D^{(d)}_{\bf p}|\Psi_0\rangle  = - 
\frac{e^{2i\theta_{i,j'}}}{\sqrt{d(d-2) + 1}} \ , \label{kvadfas} \end{equation}

\noindent and the symmetry 

\begin{equation} U_P^{(d-2)}\otimes {\bf 1}_d|\Psi_0\rangle = |\Psi_0\rangle 
\ . \end{equation}

\noindent Looking carefully at the Scott--Grassl conjectures 
\cite{Scott, Andrew} we find that they say that a SIC fiducial with this 
symmetry always exists in dimensions of the form $d(d-2)$, so this looks like 
a SIC. 

It remains to arrange that 

\begin{equation} |\langle \Psi_0|D^{(d-2)}_{\bf p}\otimes D^{(d)}_{\bf p}
|\Psi_0\rangle |^2 = \frac{1}{d(d-2) + 1} \ . \label{mal} \end{equation}

\noindent This is the hard part. However, it is at least consistent with our observation 
(in Section 3) that the Grassmannian of $(d+1)/2$-planes in ${\bf C}^d$ contains 
a Weyl--Heisenberg multiplet of planes at constant mutual chordal distance, 
see eq. (\ref{chordal}). When we change the dimension to $d-2$ and then 
factor in an extra Hilbert space of dimension $d$, it implies that can 
create a Weyl--Heisenberg multiplet consisting of $(d-2)^2$ equidistant 
subspaces of dimension $d(d-1)/2$ sitting in ${\bf C}^{d(d-2)}$. Each of 
them contains an ETF, and the total number of vectors 
is $d^2(d-2)^2$, just right for a SIC. 

The reader can see that our story exists only in bits and pieces that do not 
quite hang together. It is being improved \cite{BS}. At first it was 
told backwards \cite{aligned}. For the 22 cases where numerical 
solutions were available (or were made available by Andrew Scott) in the 
higher dimension, it was found that for every SIC in dimension $d$ one of the SICs 
in dimension $d(d-2)$ is {\it aligned} to it in the sense that it has the property 
described by eq. (\ref{kvadfas}). It was then proved that this property implies the 
existence of embedded ETFs according to the pattern we just discussed---except that 
the proof in the even dimensional case was given later \cite{AD} due to the 
complications that we have ignored here. Notice that, given that the 
aligned higher dimensional SIC fiducial vector contains only $2d(d-2)$ real numbers 
to be solved for, the number of overlap phases that are known `from below' is quite 
significant. This observation was used to obtain the solution in $d = 323 = 
19\cdot 17$ in exact form \cite{GS}. The upshot is that we know 23 instances where 
squared SIC overlap phases are helpful when we try to climb from one rung to another 
on some ladder---but we still cannot do it in an effortless manner. 

But squared SIC overlap phases also provide a concrete bridge from the SIC problem 
to the Stark conjectures \cite{Stark}. The latter were proposed in 1976, and their 
proofs would constitute a significant advance towards the solution of Hilbert's 12th. 
A little bit more precisely: It was shown by Kopp \cite{Kopp} that in some, and 
conjecturally all, prime dimensions equal to 2 modulo 3 a Galois transformation 
of the base field turns the squared SIC phases into Stark units. This has now been 
elucidated somewhat further \cite{Salamon}, and provided that the restriction to 
special choices of $d$ can be removed it seems to add credibility to our program. 

Still it may seem that we have been ignoring the harsh realities of number theory. 
They tell us that the degrees of the ray class fields rise very quickly as we 
ascend the ladders. Going from eq. (\ref{kvadfas}) to eq. (\ref{mal}) will be 
difficult. But to get to heaven it is enough 
to climb one ladder, and it could be that it would be easier to do so on a special one. 
A candidate is perhaps the ladder starting at $d = 5$, because the SIC fiducials 
appearing on its first three rungs can be written down exactly in a remarkably simple way 
\cite{SiSi}. Some of the reasons why this works continue to hold throughout the 
entire $D_0 = 3$ tower, and have to do with the way the dimensions appearing in 
the sequence decompose into primes once we are above the second rung \cite{BMun}, 
and with the fact that SIC symmetries are especially transparent in prime dimensions 
equal to 1 modulo 3 \cite{Marcus}. It also has to do with the `decoupling' 
phenomenon first observed in dimension $d = 323$ \cite{GS}, according to which 
that SIC fiducial vector can be constructed using a fairly small subfield of 
the ray class field, the cyclotomic numbers entering the displacement operators 
providing the rest. 

\vspace{10mm}

{\bf 5. The fifth section}

\vspace{5mm}

\noindent Is it likely that the SIC problem will have a happy end, in the sense that 
it will prove important for the {\sl Foundations of Physics}? I think so. The QBist 
approach to the foundations of quantum mechanics certainly suggests it \cite{Chris, 
qplex, QBism}. The number theoretical angle suggests additional arguments.  
SICs force us to pay attention to the nature of the numbers that are being used 
in quantum physics, and this shows that quantum mechanics knows more about discrete 
structures inside the continuum than one might think \cite{Erwin}. There have been 
many attempts to build up physical theory from discreteness. It may be more interesting 
to concentrate on things which, in fact, are discrete in existing theory and try 
to use them as primary concepts. (Yes, this has been said before \cite{Penrose}.) 
It is also striking to the eye that SICs 
arrange Hilbert space dimensions into ordered sequences. When the representation 
theory of Lie groups was worked out for the first time, sequences of dimensions 
such as $3, 8, 10, \dots$ must have seemed rather divorced from reality. They are 
not, as we were taught by the originators of the quark model. Because of the 
uncertain state that the SIC problem is presently in we must let the matter rest 
here, but more things may come. 

\vspace{10mm}

\noindent \underline{Acknowledgements}: I thank my students for collaboration 
and Andrei Khrennikov for the opportunity to present the SIC problem at the 
V\"axj\"o meetings, where so many interesting discussions about SICs have happened.

\end{document}